\documentclass[aps,prd,twocolumn,superscriptaddress,longbibliography]{revtex4-2}  % for review and submission
\setcitestyle{super}
\usepackage{graphicx}  % needed for figures
\usepackage{dcolumn}   % needed for some tables
\usepackage{bm}        % for math
\usepackage{amssymb}   % for math

\usepackage{physics}
\usepackage{xcolor}
\usepackage[normalem]{ulem}

\usepackage{url}\usepackage{hyperref}
\hypersetup{colorlinks,linkcolor={blue!55!black},citecolor={red!50!black},urlcolor={blue!45!black},breaklinks=true}

\def\6{{\langle}}
\def\9{{\rangle}}

\usepackage{scalerel}
\usepackage{tikz}
\usetikzlibrary{svg.path}
\definecolor{orcidlogocol}{HTML}{A6CE39}
\tikzset{
	orcidlogo/.pic={
		\fill[orcidlogocol] svg{M256,128c0,70.7-57.3,128-128,128C57.3,256,0,198.7,0,128C0,57.3,57.3,0,128,0C198.7,0,256,57.3,256,128z};
		\fill[white] svg{M86.3,186.2H70.9V79.1h15.4v48.4V186.2z}
		svg{M108.9,79.1h41.6c39.6,0,57,28.3,57,53.6c0,27.5-21.5,53.6-56.8,53.6h-41.8V79.1z M124.3,172.4h24.5c34.9,0,42.9-26.5,42.9-39.7c0-21.5-13.7-39.7-43.7-39.7h-23.7V172.4z}
		svg{M88.7,56.8c0,5.5-4.5,10.1-10.1,10.1c-5.6,0-10.1-4.6-10.1-10.1c0-5.6,4.5-10.1,10.1-10.1C84.2,46.7,88.7,51.3,88.7,56.8z};
	}
}

\newcommand\orcidlink[1]{\href{https://orcid.org/#1}{\mbox{\scalerel*{
				\begin{tikzpicture}[yscale=-1,transform shape]
					\pic{orcidlogo};
			\end{tikzpicture}}{X}}}}

\def\etal{\textit{et al.}}

\begin{document}
%\preprint{APS/123-QED}

\title{Spatiotemporal entanglement of the vacuum}

\author {Pravin Kumar Dahal\,\orcidlink{0000-0003-3082-7853}}
\email{pravin.dahal@csiro.au}
\affiliation{Commonwealth Scientific and Industrial Research Organisation (CSIRO), Clayton, VIC 3168, Australia}

\author{Kieran Hymas\,\orcidlink{0000-0003-1761-4298}}
\affiliation{Commonwealth Scientific and Industrial Research Organisation (CSIRO), Clayton, VIC 3168, Australia}

\date{\today}

\begin{abstract}

We demonstrate that the future and left Rindler wedges of Minkowski spacetime are entangled, leading to the Unruh effect. Similarly, the past and right Rindler wedges are also entangled. We propose a protocol to extract this entanglement using two two-state detectors located in the past and right Rindler wedges. By scaling the detector transition frequencies inversely with Minkowski time, entanglement from the quantum field is transferred to the detectors, suggesting they may support quantum teleportation via the vacuum. Our protocol can be implemented with current quantum systems, such as flux-tunable transmon qubits. This research provides new insights into the entanglement properties of spacetime and hints at practical applications for secure quantum information transfer using the vacuum state of a quantum field.

\end{abstract}

\maketitle

The Unruh effect is one of the simplest and most notable predictions of quantum field theory in curved spacetime~\cite{birrell1982,parker2009}.  It posits that a uniformly accelerating observer in Minkowski spacetime will perceive the inertial vacuum as a thermal bath of particles with a temperature proportional to their acceleration~\cite{unruh1976,crispino2008,ginzburg1987,sciama1981}. Despite myriad proposals~\cite{Martin-Martinez2010,ispirian2012,carballo2019unruh,sudhir2021unruh,arrechea2021inversion,quach2021}, direct detection of the Unruh effect currently remains elusive, as prohibitively large accelerations are typically required to observe even a modest Unruh temperature e.g. $a = 10^{20}$ m/s$^2$ for just $T=1$ K~\cite{davies1974}.

\begin{figure}[!t]
    \centering
    \includegraphics[width=0.35\textwidth]{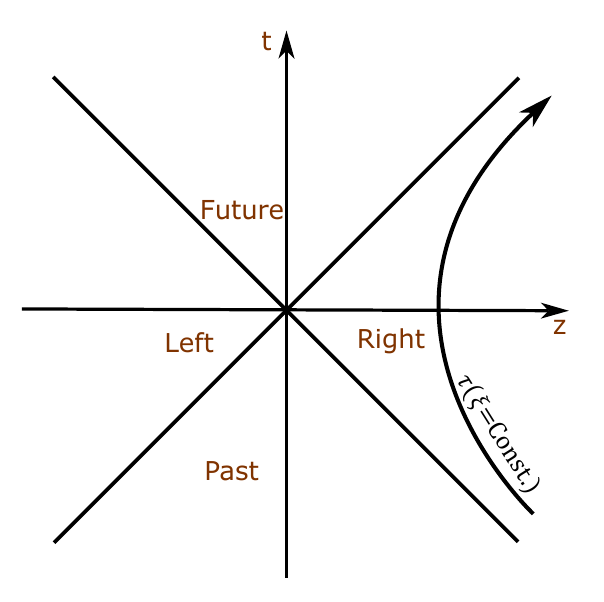}
    \caption{Minkowski spacetime split into four wedges by the null horizons corresponding to $U=0$ and $V=0$. Worldlines of constant $\xi$ are hyperbolic trajectories. A uniformly accelerated observer traces a hyperbolic trajectory of $\xi= 0$.}
    \label{fig:1}
\end{figure}

The Unruh effect originates from entanglement of  quantum field modes that pervade causally disconnected regions of Minkowski spacetime~\cite{Fuentes-Schuller2004}. An observer accelerating uniformly with $a$---known as a Rindler observer~\cite{MTW:73}---is restricted to the left or right sector bounded by the null horizons centered at some point in spacetime (Figure~\ref{fig:1}). Since the observer cannot access field modes in the opposing sector, these unobservable degrees of freedom are traced out in all local measurements of the field. The partial tracing of a globally entangled state leads to the perception of a non-empty, thermal vacuum with temperature $T$. The time-like version of the Unruh effect analogously arises from an inertial observer interacting with a quantum field in the past or future sectors of Minkowski spacetime, but whose modes are entangled across both Rindler wedges~\cite{olson2011,higuchi2017}.

Previous studies~\cite{higuchi2017,ueda2021} have demonstrated entanglement between the right-moving modes in the past-left (or future-right) Rindler wedges. Similarly, the presence of entanglement between the left-moving modes in the past-right (or future-left) wedges has also been established. However, entanglement between the right-moving modes in the past-left (or future-right) wedges does not imply that the entire past-left (or future-right) wedges are entangled—and, in fact, they are not. Consequently, the complete set of modes in the past-left (or future-right) wedges cannot constitute the full Minkowski modes. In contrast, although entanglement between the left-moving modes in the past-right (or future-left) wedges does not necessarily guarantee that the entire past-right (or future-left) wedges are entangled, they indeed are. In this work, we demonstrate that the Minkowski vacuum can be expressed as a tensor product of the past-right and future-left Rindler wedges, a result that has not yet been shown.

Entanglement harvesting—the process of extracting vacuum entanglement onto a bipartite quantum system through local interactions—has been extensively studied for spacelike-separated detectors located in the left and right Rindler wedges~\cite{reznik2000,reznik2005}. Similarly, entanglement harvesting onto timelike-separated detectors in the future and past wedges has been investigated in Ref.~\cite{olson2012,ralph2014}. These protocols require operating detectors at specific spacetime regions that are spacelike and timelike-separated, respectively. Here, we consider the spatiotemporal entanglement between mode expansions of a massive scalar field in the past and right, as well as in the left and future Rindler sectors. In addition to recovering the Unruh effect, we also propose a two detector protocol to extract entanglement from these two sectors of the Minkowski vacuum~\cite{olson2012,reznik2005,Martin-Martinez:2015,koga2018,Ng2018}. Our work enriches the Unruh literature by uncovering an additional channel to harvest vacuum entanglement between null-separated observers. Moreover, our work utilizes a conventional teleportation protocol~\cite{lin2015,landulfo2009} with accelerated (or equivalent to accelerated) partners featuring qubit degrees of freedom and entanglement resources shared between observers. This differentiates our approach from other recent proposals, such as those employing Unruh-Dewitt detectors in relative motion~\cite{koga2018} or establishing quantum teleportation of continuous variables via vacuum entanglement~\cite{foo2020}. We thus present a unique proposal for verifying past-right entanglement of Minkowski wedges that provides an alternative resource for vacuum quantum teleportation.

An Unruh–DeWitt detector with uniform acceleration $a$ corresponds to a stationary detector in the right Rindler wedge that couples to the Rindler-time positive-frequency field modes. We stress that in the future–past ‘timelike Unruh’ setting, the parameter $a$ is typically introduced as a conformal-time scaling and can be probed with stationary, time-dependent detectors. In contrast, our past-right (equivalently future–left) analysis uses a standard right-wedge Rindler description in which $a$ retains its usual role as a proper acceleration for the idealized right-wedge detector. Importantly, we highlight within our experimental proposal how this coupling can be faithfully emulated without physically accelerating the device. Since our protocol depends only on the temporal modulation of the energy levels of two %space-like separated 
detectors located at specific points in spacetime, it may be implemented with current state-of-the-art quantum systems e.g. flux-tunable transmon qubits~\cite{koch2007,zhu2021}. By transferring entanglement from the quantum field to the detectors, our protocol may act as a secure mechanism to send information~\cite{barrett2005} facilitated by vacuum entanglement.

We work with the natural system of units where $\hbar=c=1$. The Minkowski metric with signature $(-,+,+,+)$ for 3+1 spacetime is
\begin{equation}
    ds^2= -dU dV+ dx^2+dy^2 \label{eq:1}
\end{equation}
where $U = t + z$ and $V = t - z$ are light cone coordinates that describe incoming and outgoing null rays, respectively. The null rays provide a natural subdivision of Minkowski spacetime into left, right, future and past Rindler sectors where the transformation to Rindler coordinates conveniently models a uniformly accelerating trajectory in each wedge~\cite{rindler1966}. In the future sector, that is, the expanding degenerate Kasner universe, the transformation
\begin{equation}
    t= a^{-1} e^{a\chi} \cosh{a\zeta}, \quad z= a^{-1} e^{a\chi} \sinh{a\zeta}, \label{eq:2}
\end{equation}
yields the metric
\begin{equation}
    ds^2= e^{2 a \chi} (-d\chi^2+ d\zeta^2)+ dx^2+ dy^2. \label{eq:3}
\end{equation}
Substituting the light cone coordinates $u= \chi+\zeta$ and $v= \chi-\zeta$, results in the transcendental coordinate relation
\begin{equation}
    U = a^{-1} e^{au}, \label{eq:4}
\end{equation}
underpinning the Unruh effect~\cite{barcelo2011}. For completeness, the transformations in the remaining sectors are tabulated in Supplementary Table 1.

Here we consider a massive non-interacting scalar field $\phi(x)$ minimally coupled to spacetime curvature. The field evolves according to
\begin{equation}
\nabla_{\mu} \nabla^{\mu} \phi - m^2 \phi = \frac{1}{\sqrt{-g}} \partial_{\mu} \left( \sqrt{-g} g^{\mu \nu} \partial_{\nu} \phi \right) - m^2 \phi = 0, \label{eq:6}
\end{equation}
where $\nabla^\mu$ is the covariant derivative and we have assumed the Einstein summation convention where repeated Greek indices are summed from $0$ to $3$. In the future Rindler sector, $\phi(x)$ can be expanded in normalized frequency modes in the direction of acceleration (taken here to be $z$)~\cite{gerlach1988}. The positive frequency solutions are
\begin{equation}
    v_{\omega k_\perp}^F= -i \frac{e^{-i\omega\zeta+ i \mathbf{k_\perp}\cdot \mathbf{x_\perp}}}{2\pi\sqrt{4 a \sinh(\pi\omega/a)}} J_{-i\omega/a}(\kappa e^{a\chi}/a), \label{eq:7}
\end{equation}
where $\mathbf{k}_{\perp}$ are wavevectors in the transversal $\mathbf{x}_{\perp} = \left( x, y\right)$ direction and $J_{-i\omega/a}(\kappa e^{a\chi}/a)$ are Bessel functions of the first kind. The mode expansion in the left Rindler sector follows similarly~\cite{takagi1986, crispino2008} to yield the positive frequency solutions $v_{\omega k_\perp}^L$ reported explicitly in Supplementary Information Sec.~IB.

The classical field is readily promoted to a quantum field $\hat{\phi}(x)$ (with $\hbar=1$) via canonical quantization and is written as a linear superposition of all independent incoming and outgoing wave solutions~\cite{crispino2008}. Usually, the scalar field $\hat{\phi}(x)$ is written as a sum over modes in the left and right Rindler sectors since these modes form a complete set of linearly independent wave solutions in 3+1 spacetime. Using the same argument, here we write $\hat{\phi}(x)$ as the sum over modes in the left and future Rindler sectors
\begin{multline}
    \hat\phi (x)= \int_0^\infty d\omega\int d^2k_\perp \bigg(\hat a_{\omega k_\perp}^F v_{\omega k_\perp}^F+ \hat a_{\omega k_\perp}^{F\dagger} v_{\omega k_\perp}^{F *}\\
    + \hat a_{\omega k_\perp}^L v_{\omega k_\perp}^L+ \hat a_{\omega k_\perp}^{L\dagger} v_{\omega k_\perp}^{L *}\bigg). \label{eq:fex}
\end{multline}
We demonstrate in Supplementary Information Sec.~IC that the left and future solutions are complete, in the sense that the sum of their mode expansions spans the entire Minkowski spacetime. Thus, Eq.~\eqref{eq:fex} provides another valid mode expansion of the field in Minkowski spacetime~\cite{fulling1974} in addition to the familiar mode expansion in terms of plane waves~\cite{crispino2008,takagi1986,carroll2004}. This completeness would not hold if only the right-moving modes of the future and left Rindler wedges were included. Likewise, sum of mode expansions of the future and right wedges (or past and left wedges) does not constitute alternate mode expansion of Minkowski spacetime.

Using the completeness of each basis, we expand the future and left modes in plane waves
\begin{widetext}
\begin{equation}
    \begin{aligned}
        v_{\omega k_\perp}^F=& \int_{-\infty}^\infty \frac{dk_z}{\sqrt{4\pi k_0}} \left(\alpha_{\omega k_z k_\perp}^F e^{-i k_0 t+ i k_z z}+ \beta_{\omega k_z k_\perp}^F e^{i k_0 t- i k_z z}\right) \frac{e^{i \mathbf{k_\perp}\cdot \mathbf{x_\perp}}}{2\pi},\\
        v_{\omega k_\perp}^L=& \int_{-\infty}^\infty \frac{dk_z}{\sqrt{4\pi k_0}} \left(\alpha_{\omega k_z k_\perp}^L e^{-i k_0 t+ i k_z z}+ \beta_{\omega k_z k_\perp}^L e^{i k_0 t- i k_z z}\right) \frac{e^{i \mathbf{k_\perp}\cdot \mathbf{x_\perp}}}{2\pi}, \label{eq:fexp}
    \end{aligned}
\end{equation}
\end{widetext}
where $\alpha_{\omega k_z k_\perp}^F$, $\beta_{\omega k_z k_\perp}^F$, $\alpha_{\omega k_z k_\perp}^L$, $\beta_{\omega k_z k_\perp}^L$ are complex Bogoliubov coefficients independent of coordinates. In writing these expressions, we have taken into account that the Bogoliubov coefficients between modes with different $\mathbf{k}_\perp$ are zero~\cite{crispino2008}. The Bogoliubov coefficients are explicitly calculated in Supplementary Information Sec.~ID, from which we infer the following relations
\begin{equation}
\begin{aligned}
    \beta_{\omega k_z k_\perp}^F &= - e^{-\pi\omega/a} \alpha_{\omega k_z k_\perp}^{L *}\\[4mm]
    \quad \beta_{\omega k_z k_\perp}^L &= - e^{-\pi\omega/a} \alpha_{\omega k_z k_\perp}^{F *}. \label{eq:bcr}
\end{aligned}
\end{equation}
Substituting these relations into Eq.~\eqref{eq:fexp} leads to the purely positive frequency modes in Minkowski spacetime
\begin{equation}
    \begin{aligned}
        w_{-\omega k_\perp}=& \frac{v_{\omega k_\perp}^F+ e^{-\pi\omega/a} v_{\omega -k_\perp}^{L*}}{\sqrt{1- e^{-2\pi\omega/a}}}\\[2mm]
        w_{\omega k_\perp}=& \frac{v_{\omega k_\perp}^L+ e^{-\pi\omega/a} v_{\omega -k_\perp}^{R*}}{\sqrt{1- e^{-2\pi\omega/a}}} \label{eq:ppm}
    \end{aligned}
\end{equation}
where
\begin{multline}
    w_{\pm\omega k_\perp}= \int_{-\infty}^{\infty} \frac{dk_z}{\sqrt{8 a} \pi k_0} \left(\frac{k_z+k_0}{k_z-k_0}\right)^{\pm i\omega/2a} \\ \times e^{- i k_0 t+ i k_z z} \frac{e^{i \mathbf{k_\perp}\cdot \mathbf{x_\perp}}}{2\pi}. \label{eq:mmo}
\end{multline}
Expanding the field $\hat{\phi}(x)$ as a linear combination of these modes and their complex conjugates, it is evident that the operators $\hat a_{\omega k_\perp}^F- e^{-\pi\omega/a} \hat a_{\omega -k_\perp}^{L\dagger}$ and $\hat a_{\omega k_\perp}^L- e^{-\pi\omega/a} \hat a_{\omega -k_\perp}^{F\dagger}$ both annihilate the Minkowski vacuum state $\ket{0_M}$
\begin{equation}
    \begin{aligned}
        \left(\hat a_{\omega k_\perp}^F- e^{-\pi\omega/a} \hat a_{\omega -k_\perp}^{L\dagger}\right)|0_M\rangle &= 0\\[2mm]
        \left(\hat a_{\omega k_\perp}^L- e^{-\pi\omega/a} \hat a_{\omega -k_\perp}^{F\dagger}\right)|0_M\rangle &= 0. \label{eq:anop}
    \end{aligned}
\end{equation}
The Rindler vacuum state $\ket{0_R}$ is defined as the state of no particles corresponding to the Rindler metric, and obeys the identities
\begin{equation}
    \hat a_{\omega k_\perp}^F|0_R\rangle= \hat a_{\omega k_\perp}^L|0_R\rangle= 0 \quad \textrm{for all} \quad \omega, k_\perp.
\end{equation}

Assuming the frequencies $\omega$ faithfully go over to a set of discrete frequencies $\omega_i$ as in the typical box quantization scheme~\cite{crispino2008}, Eq.~\eqref{eq:anop} implies that
\begin{equation}
    \left(\hat a_{\omega_i k_\perp}^{F\dagger}\hat a_{\omega_i k_\perp}^F- \hat a_{\omega_i, -k_\perp}^{L\dagger} \hat a_{\omega_i, -k_\perp}^{L}\right)|0_M\rangle= 0.
\end{equation}
Thus, for each $\omega_i$, an equal number of left and future Rindler modes exist. Consequently,
\begin{equation}
\begin{aligned}
    |0_M\rangle &\propto \prod_i\sum_{n_i=0}^\infty\int dk_\perp\frac{K_{n_i}}{n_i!} \left(\hat a_{\omega_i k_\perp}^{F\dagger} \hat a_{\omega_i k_\perp}^{L\dagger}\right)^{n_i} |0_R\rangle \\[4mm]
    &\propto \prod_i\sum_{n_i=0}^\infty\int dk_\perp K_{n_i}|n_i,F\rangle\otimes |n_i,L\rangle, \label{eq:ent2}
\end{aligned}
\end{equation}
where $K_{n_i}$ is a proportionality factor determined by the discretised analogue of Eq.~\eqref{eq:anop} and the normalization condition on $\ket{0_M}$. From Eq. (\ref{eq:ent2}), we have demonstrated that the Minkowski vacuum decomposes into an entangled superposition of $|n_i,F\rangle$ and $|n_i,L\rangle$ Rindler states, each describing $n_i$ particles with Rindler energy $\omega_i$ in the future and left wedges, respectively. After tracing out unobserved modes in the future Rindler sector, a Rindler observer in the left wedge perceives the Minkowski vacuum as a thermal state with temperature $T = a/2\pi$. Proceeding in a similar manner, one finds that the scalar field is also entangled in the past and right Rindler sectors of Minkowski spacetime.

We now describe a protocol to extract the field entanglement via bipartite entanglement of two Unruh-Dewitt detectors with level spacing $E$, in causally disconnected regions of Minkowski spacetime~\cite{reznik2005}. The two Unruh-Dewitt detectors, initially unentangled at $t=-\infty$, are confined to the past and right Rindler sectors of an inertial observer and interact locally with the vacuum state of the scalar field for some finite interval of Rindler coordinate time. The detectors select out a preferred moment in spacetime---in this case the origin $(t,z)=(0,0)$---and entanglement between the field modes is appropriately redistributed to the detectors as $t \rightarrow \infty$. We demonstrate that detectors in the past and right Rindler sectors, whose spacetime interactions are symmetric across the origin, can transform the past-right vacuum entanglement into constant-time bipartite entanglement at $t=\infty$. In realistic experiments, the primary interactions between detectors and quantum fields are expected to arise from the massless vacuum electromagnetic field. Since the results obtained for a massive field reduce to those for a massless field in the limit $m \rightarrow 0$, our conclusions remain valid in the massless case as well.

The detectors in the past and right sectors move along trajectories parametrized by $x=y=z=0, t=-a^{-1} e^{a\chi}$ and $x=y=0, z= a^{-1} \cosh{a\tau}, t= a^{-1} \sinh{a\tau}$, respectively~\cite{olson2012}. The dynamics in the Rindler coordinate time $\tau$ of the right detector are generated by the Hamiltonian $H_R(\tau) = H_0 + H_I(\tau)$ where $H_0 = E |1\rangle \langle 1|$ and $H_I(\tau)= m(\tau) \hat\phi(z(\tau)) |1\rangle \langle 0|+ \text{H.c.}$ describes the interaction of the detector monopole $m(\tau)= e^{iH_0\tau} m(0) e^{-iH_0\tau}$ with the field $\hat\phi$~\cite{crispino2008}. The conformal time dynamics of the past detector are generated by a similar Hamiltonian $H_P(\chi) = H_0 + H_I(\chi)$. Subtlety, passing from the conformal time $\chi$ of the past detector to Minkowski time $t$ reveals that $H_0$ goes over to $-H_0/ a t$ (Supplementary Information Sec. II). A detector moving along the trajectory $x=y=z=0, t=-a^{-1} e^{a\chi}$ in the past Rindler sector is thus equivalent to a stationary detector whose level spacing $E$ is detuned in Minkowski time~\cite{olson2011}. Flux-tunable transmon qubits are well-suited for realizing such a detector in practice~\cite{krantz2019}.

A positive detector negativity $\mathcal{N}(\rho)$ at $t=\infty$ is indicative of entanglement between the detectors, a necessary and sufficient condition for the non-separability of the two-detector density operator $\rho = |\psi\rangle \langle \psi|$~\cite{peres1996,horodecki1996}. We calculate the state of the two detectors at $t=\infty$ to second order in the field-detector interaction Hamiltonian $H_I$ using perturbation theory in the Rindler coordinate time of the detectors. We account for the adiabatic switching on and off of the detectors with window functions ${\cal E}_P(\chi)$ and ${\cal E}_R(\tau)$ which are assumed to be piecewise smooth functions in the Rindler coordinate time of each detector.

To second order in the perturbation, the two-detector state at $t=\infty$ takes the form
\begin{equation}
    |\psi\rangle\approx |1-C\rangle |00\rangle - i |A_R\rangle|01\rangle- i |A_P\rangle|10\rangle-|X\rangle|11\rangle,
\end{equation}
where $|C\rangle$, $\ket{A_R}$, $\ket{A_P}$ and $\ket{X}$ are integrals of the field operators over Rindler coordinate time applied to the Minkowski vacuum state defined explicitly in Supplementary Information Sec.~II. The negativity ${\cal N}(\rho)$ approximates to~\cite{reznik2005,olson2012}
\begin{equation}
    {\cal N}(\rho)\approx \frac{|\langle0_M|X\rangle|- \sqrt{\langle A_R|A_R\rangle\langle A_P|A_P\rangle}}{N},
\end{equation}
where $N$ accounts for normalization of the two-detector reduced density matrix. If the lineshapes of ${\cal E}_P(\chi)$ and ${\cal E}_R(\tau)$ are equal then $\langle A_R|A_R\rangle= \langle A_P|A_P\rangle$ and the non-separability condition amounts to
\begin{widetext}
    \begin{equation} \label{eq:ixia}
    \bigg|\int d\tau\int d\chi {\cal E}_R(\tau) {\cal E}_P(\chi) e^{a\chi} e^{-iE(\tau-\chi)} \langle 0_M|\hat\phi(\tau) \hat\phi(\chi)|0_M\rangle\bigg|> \bigg|\int d\tau\int d\tau' {\cal E}_R(\tau) {\cal E}_R(\tau') e^{-iE(\tau-\tau')} \langle 0_M|\hat\phi(\tau') \hat\phi(\tau)|0_M\rangle\bigg|
    \end{equation}
\end{widetext}
which is compactly expressed as $I_X>I_A$.

For the purpose of calculation, we choose the window functions ${\cal E}_P(\chi)= e^{-\chi^2}$ and ${\cal E}_R(\tau)= e^{-\tau^2}$ and evaluate the right-hand side of Eq.~(\ref{eq:ixia}) with the dimensionless parameters $E=1$ and $a=2$ to obtain $I_A= 1.273$, consistent with previous results~\cite{olson2012}. In Supplementary Information Sec.~II we evaluate $I_X \approx 2.179$ using the same parameters. Since $I_X > I_A$, the positive two-detector negativity indicates that the detectors are entangled at $t=\infty$. Symmetric translation of the Gaussian window functions about the origin $(t,z)=(0,0)$, such that their peaks in the respective past and right Rindler wedges remain null-separated, amounts to a new pair of window functions that leave the sign of the negativity invariant. However, to achieve the same degree of entanglement between the detectors requires that they interact with the field for longer spacetime intervals i.e. that the full widths at half maximum of the window functions ${\cal E}_{P/R}$ are broadened proportional to their displacement~\cite{olson2012}.

\begin{figure}[t]
    \centering
    \includegraphics[width=0.95\linewidth]{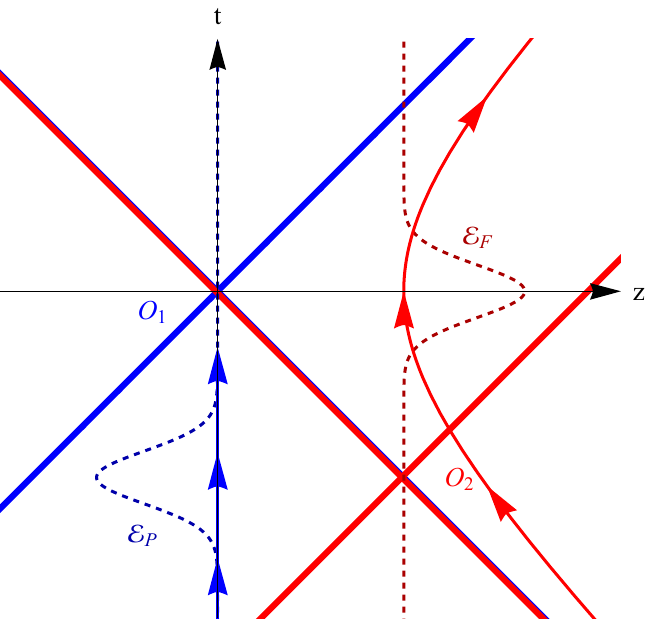}
    \caption{Schematic representation of the light cones of null separated observers $O_1$ (solid blue) and $O_2$ (solid red) in Minkowski spacetime. The trajectories of past and right detectors in the reference frame of $O_1$ are shown as blue and red curves, respectively, with arrows indicating forward evolution in time. In this frame, past detector evolves asymptotically towards the origin, whilst right detector follows a hyperbolic path. The window functions ${\cal E}$ are shown as dashed lines as function of each observers coordinate time. The peaks in ${\cal E}$ indicate regions of spacetime where $O_i$'s detector is active. In this schematic, note that the detector of $O_2$ is active in the future light cone of $O_2$, corresponding to the right Rindler wedge of $O_1$.}
    \label{fig:2}
\end{figure}

Until now, we have considered a uniformly accelerated Unruh–DeWitt detector in the right Rindler sector of an observer $O_1$, and in this case the parameter $a$ has its usual interpretation as the detector’s proper-acceleration. Importantly, physical acceleration is not required for the experimental realization of the past-right vacuum entanglement. We now present an operationally equivalent, experimentally accessible alternative in which both detectors remain stationary but their level spacings are engineered in time so as to reproduce the same effective interaction (within a specific interaction window) with the relevant field modes. Given the initial observer $O_1$, we specify a null separated observer $O_2$ whose future light cone coincides with the right and future sectors of $O_1$ (Figure 2). For a time interval determined by the separation between $O_1$ and $O_2$, a second stationary detector in the future light cone of $O_2$---spacelike-separated from $O_1$---interacts with the vacuum fluctuations in the right Rindler wedge of $O_1$. Transforming from the conformal time of this second detector to Minkowski time reveals a similar energy level rescaling, $H_0 \rightarrow H_0/at$, analogous to that of the detector in the past light cone of $O_1$. Below, we identify the limit in which this energy detuning aligns with that of a physically accelerating observer in the right Rindler wedge of $O_1$.

The trajectory of a future detector $x=y=z_2=0,\, t_2=a^{-1} e^{a\chi_2}$ corresponding to the observer $O_2$ can be achieved by scaling its energy level. This trajectory on the future wedge of observer $O_2$ is equivalent to the trajectory of a uniformly accelerating detector $x=y=0,\, z_1= a^{-1} \cosh{a\tau_1},\, t_1= a^{-1} \sinh{a\tau_1}$ on the right wedge of observer $O_1$ if $a\chi_2= a\tau_1\ll 1$. Thus, for some specific $a$, the window functions of the detectors should be adjusted such that $a\chi_2= a\tau_1\ll 1$ is satisfied. This means that, depending on the energy scaling of detectors, their interaction with a field is limited to the time ${\cal O} \left(\tau_1,\chi_2\right)\ll 1/a$. We thus choose window functions ${\cal E}_R(\tau_1)= e^{-\tau_1^2/\sigma}, \, {\cal E}_F(\chi_2)= e^{-\chi_2^2/\sigma}$, such that their values drop by at least an order of magnitude approximately after time ${\cal O} \left(\tau_1,\chi_2\right)\ll 1/a$. We then evaluate the negativity of the two-detector state (past-right of $O_1$ or future-left of $O_2$) numerically. We obtain the result that the detector negativity along this new trajectory is positive indicating that detector entanglement is maintained at $t = \infty$ for $E=25$ GHz, $a=50$ GHz, and $\sqrt{\sigma} = 1.32$ ps, which are chosen to be compatible with the above bounds. Broader window functions, such as those feasible in superconducting qubit set-ups~\cite{wang2020controllable,zhu2021}, also preserve detector entanglement. We considered the energy scaled detector on the future trajectory of $O_2$ and another energy scaled detector on the past trajectory of $O_1$. Evaluating the negativity of the two-detector state numerically with an increased width of the window function $\sqrt{\sigma} = 0.5$ ns (and all other parameters unchanged) results in a positive detector negativity thereby indicating that detectors are still entangled at $t=\infty$. This implies that bipartite entanglement can be transferred to both detectors from the entangled vacuum modes by conformally rescaling both of their excitation frequencies. 

We point out that, since spacetime points in the past and right Rindler sectors are not causally disconnected, correlations between detectors could also arise via classical communication channels. For the conclusive demonstration of bipartite detector entanglement facilitated by the quantum vacuum, it is thus imperative for any experimental implementation to suppress all classical (and all other quantum for that matter) communication channels between the detectors.

The two-detector setup described above not only lends itself to the experimental verification of our proposal, but also---by using the vacuum field to generate the entanglement resource---opens up the possibility of secure quantum communication between the two null-separated observers. We consider a quantum teleportation protocol~\cite{bennett1992} in which the entanglement resource is between detectors interacting in the past and right Rindler sectors of $O_1$. According to the protocol outlined above, the detectors become entangled at some point in the distant future after their interaction with the vacuum at preordained spacetime intervals symmetric about $(t, z) = (0, 0)$. At that time, $O_1$ can perform a joint measurement on detector 1 and the qubit to be teleported. By transmitting the results through a classical channel, $O_2$ can apply appropriate unitary operations to the second detector, thereby reconstructing the state of the original qubit.

We have shown that the future and left Rindler wedges, as well as the past and right Rindler wedges of Minkowski spacetime, are entangled---a structure that recovers the Unruh effect upon tracing out unobserved modes from a given sector. To harness this entanglement, we proposed a protocol using two two-level detectors positioned in the past and right Rindler wedges of an observer. Remarkably, this entanglement can be extracted and potentially used as a resource for secure quantum information transfer. The protocol is compatible with current quantum technologies, such as flux-tunable transmon qubits, making experimental verification and real-world applications feasible. By revealing entanglement between distinct Rindler wedges, our work also offers new perspectives on the quantum structure of spacetime and contributes to the emerging field of relativistic quantum information~\cite{reznik2002,bradler2007}, with potential implications for quantum gravity.

\section*{Acknowledgements}
P.K.D. would like to thank Nicolas C. Menicucci, Daniel R. Terno, Timothy C. Ralph, Eduardo Martin-Martinez and Yaniv Kurman for useful comments. The work of P.K.D. is supported by CSIRO Early Research Career postgraduate fellowship. K.H. acknowledges financial support from the Revolutionary Energy Storage Systems Future Science Platform.

\clearpage
\widetext
\begin{center}
\textbf{\large Supplementary Information: Spatiotemporal entanglement of the vacuum}
\end{center}

%\title{Supplementary Information: Spatiotemporal entanglement of the vacuum}

\twocolumngrid

\setcounter{equation}{0}
\setcounter{figure}{0}
\setcounter{table}{0}
\setcounter{page}{1}
\makeatletter
\renewcommand{\theequation}{S\arabic{equation}}
\renewcommand{\thefigure}{S\arabic{figure}}
\renewcommand{\bibnumfmt}[1]{[S#1]}
\renewcommand{\citenumfont}[1]{S#1}

\section{Scalar field in Rindler Spacetime}

\subsection{Rindler coordinates in each sector} \label{s1A}

\begin{table}[h]
    \centering
    \begin{tabular}{|c|cc|}
        \hline
        Sector & t & z \\
         \hline
        Future & $a^{-1} e^{a\chi} \cosh{a\zeta}$ & \quad $a^{-1} e^{a\chi} \sinh{a\zeta}$ \\
        Past & $-a^{-1} e^{a\chi} \cosh{a\zeta}$ & \quad $-a^{-1} e^{a\chi} \sinh{a\zeta}$ \\
        Left & $-a^{-1} e^{a\xi} \sinh{a\tau}$ & \quad $-a^{-1} e^{a\xi} \cosh{a\tau}$ \\
        Right & $a^{-1} e^{a\xi} \sinh{a\tau}$ & \quad $a^{-1} e^{a\xi} \cosh{a\tau}$ \\
        \hline
        \end{tabular}
    \caption{Rindler coordinate transformations for the four sectors of Minkowski spacetime.}
\end{table}

As the metric components in Rindler coordinates do not depend on $\zeta$, $\partial_\zeta$ is a timelike Killing vector of the Rindler metric. The two surface, where this Killing vector becomes null is the Killing horizon and is given by $\chi=-\infty$ or $t=\pm z$ (see Fig.~1 of the article).

\subsection{Mode expansion of a massive scalar field in each Rindler sector}

\subsubsection{Future sector}

We consider a massive non-interacting scalar field $\phi$ minimally coupled to spacetime curvature. The field evolves according to
\begin{equation}
\nabla_{\mu} \nabla^{\mu} \phi - m^2 \phi = \frac{1}{\sqrt{-g}} \partial_{\mu} \left( \sqrt{-g} g^{\mu \nu} \partial_{\nu} \phi \right) - m^2 \phi = 0, \label{eq:6s}
\end{equation}
where $\nabla^\mu$ is the covariant derivative. Expand this equation in the future Rindler sector gives
\begin{equation}
    \partial_\zeta^2\phi- \partial_i \left(g^{ij} e^{2 a \chi}\partial_j\phi\right)+ e^{2 a \chi} m^2\phi= 0.
\end{equation}
We solve this equation using the method of separation of variables. The solution corresponding to the temporal variable is $e^{\pm i\omega\zeta}$, where $\omega$ is a nonnegative constant. For nonnegative $\omega$, with respect to the (right oriented) spacelike Killing vector, it is natural to assume that negative sign of the exponent gives the positive frequency solution. Moreover, the solution in the transverse $(x,y)$ direction has the form $e^{\pm i\mathbf{k}_\perp\cdot\mathbf{x}_\perp}$.

Thus, the equation for field $\phi$ reduces to the ordinary differential equation in the variable $\chi$ only
\begin{equation}
    \left(- \frac{d^2}{d\chi^2}- e^{2a\chi}\left(k_\perp^2+ m^2\right)\right)g_{\omega k_\perp}= \omega^2 g_{\omega k_\perp},
\end{equation}
where $g_{\omega k_\perp}$ is the component of $\phi$, depending only on coordinate $\chi$.The solution of this equation is the linear combination of Bessel functions $J_{-i\omega/a}(\kappa e^{a\chi}/a)$ and $J_{i\omega/a}(\kappa e^{a\chi}/a)$, where $\kappa= k_\perp^2+ m^2$. As the physically relevant solutions tend to zero as $\chi\to +\infty$ and oscillate like $e^{\pm i\omega\chi}$ as $\chi\to -\infty$, we disregard the solution containing $J_{i\omega/a}(\kappa e^{a\chi}/a)$. We thus obtain the normalized positive frequency mode in the future Rindler sector
\begin{equation}
    v_{\omega k_\perp}^F= -i \frac{e^{-i\omega\zeta+ i \mathbf{k_\perp}\cdot \mathbf{x_\perp}}}{2\pi\sqrt{4 a \sinh(\pi\omega/a)}} J_{-i\omega/a}(\kappa e^{a\chi}/a). \label{eq:fmods}
\end{equation}

\subsubsection{Left sector}

The mode expansion in the left Rindler sector, $v_{\omega k_\perp}^L$,  are available in the literature (we refer readers to Refs.~\cite{crispino2008s,takagi1986s} for its derivation). The normalized positive frequency mode in the left Rindler sector is
\begin{equation}
    v_{\omega k_\perp}^L= \frac{e^{i\omega\tau+ i \mathbf{k_\perp}\cdot \mathbf{x_\perp}}}{\sqrt{4\pi^4 a \csch(\pi\omega/a)}} K_{i\omega/a}(\kappa e^{a\xi}/a)~,
\end{equation}
where $K_{i\omega/a}(\kappa e^{a\xi}/a)$ are modified Bessel functions.

\subsection{Completeness of the modes in future and left Rindler sectors}

The arguments for the completeness of the modes in the future and left Rindler sectors follow analogously to those for the completeness of the modes in the right and left Rindler sectors. For detailed calculations, readers are directed to Ref.~\cite{crispino2008s}. The modes $v_{\omega k_\perp}^F$ and $v_{\omega k_\perp}^L$, which vanish outside the future and left Rindler sectors respectively, can be expressed in terms of the purely positive frequency modes in Minkowski spacetime (written in Eq.~11 of the article) as
\begin{equation}
    \begin{aligned}
        v_{\omega k_\perp}^F=& \frac{w_{-\omega k_\perp}- e^{-\pi\omega/a} w_{\omega -k_\perp}^{*}}{\sqrt{1- e^{-2\pi\omega/a}}},\\
        v_{\omega k_\perp}^L=& \frac{w_{\omega k_\perp}- e^{-\pi\omega/a} w_{-\omega -k_\perp}^{*}}{\sqrt{1- e^{-2\pi\omega/a}}}. \label{eq:s6}
    \end{aligned}
\end{equation}
This relation, along with Eq.~12 of the article give the modes $v_{\omega k_\perp}^F$ and $v_{\omega k_\perp}^L$ as distributions in the whole of Minkowski spacetime. These modes form a complete set of solutions to the Klein-Gordon equation, not only in the future and left Rindler sectors but in the entire Minkowski spacetime.

Analogous calculations to those performed in the right and left Rindler wedges~\cite{crispino2008s} can be carried out to verify that the Wightman two-point function is correctly reproduced throughout Minkowski spacetime, even when using future and left Rindler modes. To begin, we use Eqs.~(12) to write the Rindler annihilation operators as
\begin{equation}
    \begin{aligned}
        \hat a_{\omega k_\perp}^F= \frac{\hat b_{-\omega k_\perp}+ e^{-\pi\omega/a} \hat b_{\omega -k_\perp}^\dagger}{\sqrt{1- e^{-2\pi\omega/a}}}~,\\
       \hat a_{\omega k_\perp}^L= \frac{\hat b_{\omega k_\perp}+ e^{-\pi\omega/a} \hat b_{-\omega -k_\perp}^\dagger}{\sqrt{1- e^{-2\pi\omega/a}}}~. 
    \end{aligned}
\end{equation}
Here $\hat b_{\pm\omega k_\perp}$ denote the Minkowski vacuum annihilation operators, which satisfy the commutation relations
\begin{equation}
    \left[\hat b_{\pm\omega k_\perp},\hat b_{\pm\omega' k_\perp'}^\dagger\right]= \delta(\omega-\omega') \delta^2(k_\perp-k_\perp')~,
\end{equation}
with all other commutators vanishing. These relations can be used to calculate the vacuum expectation values
\begin{widetext}
\begin{equation}
\begin{aligned}
\langle0_M|\hat a_{\omega k_\perp}^{F\dagger} \hat a_{\omega' k_\perp'}^F|0_M\rangle=& \langle0_M|\hat a_{\omega k_\perp}^{L\dagger} \hat a_{\omega' k_\perp'}^L|0_M\rangle= \left(e^{2\pi\omega/a}- 1\right)^{-1} \delta(\omega-\omega') \delta^2(k_\perp-k_\perp')~,\\
\langle0_M|\hat a_{\omega k_\perp}^{F} \hat a_{\omega' k_\perp'}^{F\dagger}|0_M\rangle=& \langle0_M|\hat a_{\omega k_\perp}^{L} \hat a_{\omega' k_\perp'}^{L\dagger}|0_M\rangle= \left(1- e^{-2\pi\omega/a} \right)^{-1} \delta(\omega-\omega') \delta^2(k_\perp-k_\perp')~,\\
\langle0_M|\hat a_{\omega k_\perp}^{L} \hat a_{\omega' k_\perp'}^{F}|0_M\rangle=& \langle0_M|\hat a_{\omega k_\perp}^{L\dagger} \hat a_{\omega' k_\perp'}^{F\dagger}|0_M\rangle= \left(e^{\pi\omega/a}- e^{-\pi\omega/a} \right)^{-1} \delta(\omega-\omega') \delta^2(k_\perp+k_\perp')~,
\end{aligned}
\end{equation}
with vacuum expectation values of all other products of creation and annihilation operators vanishing. We can use these relations to write the two-point function for the field given in Eq.~(7) as
\begin{multline}
    \Delta(x;x')= \int_0^\infty d\omega \int d^2k_\perp \bigg( \left(v_{\omega k_\perp}^F(x) v_{\omega k_\perp}^{F *}(x')+ v_{\omega k_\perp}^L(x) v_{\omega k_\perp}^{L *}(x') \right) \left(1- e^{-2\pi\omega/a} \right)^{-1}+\\
    \left(v_{\omega k_\perp}^{F*} (x) v_{\omega k_\perp}^{F}(x')+ v_{\omega k_\perp}^{L*}(x) v_{\omega k_\perp}^{L}(x') \right) \left( e^{2\pi\omega/a}- 1 \right)^{-1}+ 2 \left(v_{\omega k_\perp}^{F} (x) v_{\omega -k_\perp}^{L}(x')+ v_{\omega k_\perp}^{F*}(x) v_{\omega -k_\perp}^{L*}(x') \right) \left( e^{\pi\omega/a}- e^{-\pi\omega/a} \right)\\
    + 2 \left(v_{\omega k_\perp}^{L} (x) v_{\omega -k_\perp}^{F}(x')+ v_{\omega k_\perp}^{L*}(x) v_{\omega -k_\perp}^{F*}(x') \right) \left( e^{\pi\omega/a}- e^{-\pi\omega/a} \right)\bigg)~.
\end{multline}
\end{widetext}
Using Eqs.~\eqref{eq:s6}, this relation simplifies to
\begin{multline}
    \Delta(x;x')= \int_0^\infty d\omega \int d^2k_\perp \\
    \left( w_{\omega k_\perp}(x) w_{\omega k_\perp}^*(x')+ w_{-\omega k_\perp}(x) w_{-\omega k_\perp}^*(x') \right)~. \label{eq:s11}
\end{multline}
Substituting $w_{\omega k_\perp}$ from Eq.~(11) and performing the integration yields the two-point function
\begin{equation}
    \Delta(x;x')= \int \frac{dk_z d^2k_\perp}{2 k_0 (2\pi)^3} e^{i k\cdot (x-x')}~,
\end{equation}
which coincides with the two-point function in the Minkowski vacuum state. Eq.~\eqref{eq:s11} is undefined when either of the two points lies on the hyperplane $t=\pm z$. However, since the two-point function is defined as a distribution, it remains well-defined on Minkowski spacetime when smeared with test functions of compact support~\cite{crispino2008s}.

We have explicitly calculated the Bogoliubov coefficients in the future wedge below. Moreover, explicit calculations for the right wedge are available in the literature~\cite{crispino2008s}. From these results, we infer the following relations between the coefficients
\begin{equation}
\beta_{\omega k_z k_\perp}^F= \beta_{\omega k_z k_\perp}^{R}~, \quad
    \quad \alpha_{\omega k_z k_\perp}^R= \alpha_{\omega k_z k_\perp}^{F}~.
\end{equation}
As a consequence, the Minkowski modes cannot be expressed as linear combinations of modes in the future and right sectors alone. Therefore, sum total of modes in the future and right Rindler wedges could not generate complete Minkowski modes. Similarly, sum total of only left-moving modes in the past–right (or future–left) wedges is also insufficient to generate a complete set of Minkowski modes.

\subsection{Bogoliubov coefficients}

As Bogoliubov coefficients are coordinate independent, calculations of them can be simplified by taking the limit of the solutions on the future Killing horizon (see Sec.~\ref{s1A})
\begin{multline}
    v_{\omega k_\perp}^F\bigg|_{t=z}= \int_{-\infty}^\infty \frac{dk_z}{\sqrt{4\pi k_0}} \\ \left(\alpha_{\omega k_z k_\perp}^F e^{-i (k_0 - k_z) V/2}+ \beta_{\omega k_z k_\perp}^F e^{i (k_0 - k_z) V/2}\right) \frac{e^{i \mathbf{k_\perp}\cdot \mathbf{x_\perp}}}{2\pi}. \label{eq:minaps}
\end{multline}
On approaching the Killing horizon, we also have $\chi\to-\infty$ and thus, we can simplify the Bessel function using the small argument approximation there
\begin{equation}
    J_\nu(x)\approx \left(\Gamma(\nu+1)\right)^{-1} (x/2)^\nu.
\end{equation}
Thus, Eq.~\eqref{eq:fmods} approximates to
\begin{multline}
    v_{\omega k_\perp}^F\bigg|_{\chi\to-\infty}= \\
    -i \frac{e^{i \mathbf{k_\perp}\cdot \mathbf{x_\perp}}}{2\pi\sqrt{4 a \sinh(\pi\omega/a)}} \left(\frac{\kappa}{2 a}\right)^{-i\omega/a} \frac{e^{-i\omega\zeta- i \omega\chi}}{\Gamma(1- i\omega/a)}. \label{eq:rinaps}
\end{multline}
We thus substitute Eqs.~\eqref{eq:minaps} and \eqref{eq:rinaps} into Eq.~9 of the article to calculate Bogoliubov coefficients
\begin{equation}
    \begin{aligned}
        \alpha_{\omega k_z k_\perp}^F=& -\frac{i \left(\kappa/2 a\right)^{-i\omega/a} (k_0- k_z)}{4\sqrt{\pi a k_0 \sinh(\pi\omega/a)} \Gamma(1- i\omega/a)}\\
        & \int_0^\infty dV (a V)^{-i\omega/a} e^{i(k_0-k_z)V/2}\\
        =& \frac{e^{\pi\omega/2a}}{\sqrt{4 \pi a k_0 \sinh(\pi\omega/a)}} \left(\frac{k_z+k_0}{k_z-k_0}\right)^{-i\omega/2a},
    \end{aligned}
\end{equation}
\vspace{5mm}
and
\begin{equation}
    \beta_{\omega k_z k_\perp}^F
    = - \frac{e^{-\pi\omega/2a}}{\sqrt{4 \pi a k_0 \sinh(\pi\omega/a)}} \left(\frac{k_z+k_0}{k_z-k_0}\right)^{-i\omega/2a},
\end{equation}
where $\kappa= \sqrt{(k_z-k_0)(k_z+k_0)}$ is used. Explicit calculations of $\alpha_{\omega k_z k_\perp}^L$ and $\beta_{\omega k_z k_\perp}^L$ are available in the literature (we refer readers to Refs.~\cite{crispino2008s,takagi1986s}).

\hspace{5mm}

\section{Entanglement}

To see the manifestation of the timelike Unruh effect in the particle detector in the future-past (FP) coordinates, consider the Schrodinger equation for an observer in the worldline $(\eta,0)$
\begin{equation}
    i\hbar\frac{\partial\psi}{\partial \eta}= H\psi.
\end{equation}
In Minkowski coordinates, this equation becomes
\begin{equation}
    i\hbar\frac{\partial\psi}{\partial t}= \frac{H}{a t} \psi.
\end{equation}\\[1mm]
This implies that a detector at rest in the FP coordinates corresponds to a detector with energy levels scaled inversely with time in Minkowski coordinates. Details on the choice of the coupling constant of such a detector, ensuring that perturbation theory is valid, are discussed in Ref.~\cite{olson2011s}.

At $t=-\infty$, we take the state of the detectors to be $|00\rangle$. Unitary evolution of the initial quantum state in the interaction picture is given by the following relation
\begin{equation}
    |\psi\rangle= U|0_M\rangle|00\rangle,
\end{equation}
where $U$ is the time-ordered propagator
\begin{equation}
    U= \hat T \left(e^{-i\int H_I(\chi) d\chi} e^{-i\int H_I(\tau) d\tau}\right)~. \label{eq:s23}
\end{equation}
This guarantees that $U$ does not change net entanglement between the regions. In the past light cone, the Rindler time coordinate $\chi$ ranges from $-\infty$ to $\infty$, while the Minkowski time $t$ ranges from $-\infty$ to $0$. Consequently, the time integrals of the interaction Hamiltonian for a detector following the worldline $z(t)$ are related by
\begin{equation}
\int_{-\infty}^\infty H_I(\chi) d\chi= \int_{-\infty}^0 H_I[\chi(t)] \frac{d\chi}{dt}  dt~.
\end{equation}
Since the past detector interacts with the field only within the region constrained by the Gaussian window function $e^{-\chi^2}$, the interaction is confined to the interval $t \in (-\infty, 0]$, and the interaction Hamiltonian vanishes identically for $t \in (0, \infty)$. Therefore, the time integral of the interaction Hamiltonian over the full Minkowski time axis becomes
\begin{equation}
\int_{-\infty}^\infty H_I(\chi) d\chi= \int_{-\infty}^\infty H_I[\chi(t)] \frac{d\chi}{dt} dt~,
\end{equation}
where $H_I[\chi(t)] := 0$ for $t \geq 0$. For completeness, we have taken $\chi = t$ in the interval $t \in (0, \infty)$, corresponding to the continuation of the past detector along an inertial trajectory. Hence, the unitary operator~\eqref{eq:s23} describes the evolution of the two-detector system as both detectors evolve to $t=\infty$ in the future.

We work in the regime where contributions from the detector-field interaction can be treated perturbatively. To second order in perturbation theory, the unitary at $t=\infty$ takes the following form
\begin{widetext}
\begin{multline}
        U\approx \hat T \bigg(\mathbf{1}-i\int H_I(\chi) d\chi-i\int H_I(\tau) d\tau- \int d\chi \int d\tau H_I(\chi) H_I(\tau)\\
        - \frac{1}{2} \int d\chi \int d\chi' H_I(\chi) H_I(\chi') - \frac{1}{2} \int d\tau \int d\tau' H_I(\tau) H_I(\tau') \bigg).
    \end{multline}
The interaction Hamiltonians of the past and right detectors with the field have the form $H_I= m(\chi) \hat\phi(x(\chi)) |1\rangle \langle 0|+ H.c.$, $m(\chi)= e^{iH_0\chi} m(0) e^{-iH_0\chi}$ being the detector's monopole. With this interaction Hamiltonian, the state at $t=\infty$ becomes
\begin{equation}
    |\psi\rangle\approx (1-C) |00\rangle - i |A_R\rangle|01\rangle- i |A_P\rangle|10\rangle-|X\rangle|11\rangle,
\end{equation}
where we have defined the following (unnormalized) field states to simplify the notation
\begin{equation}
\begin{aligned}
    |C\rangle=& \int d\tau\int d\tau' {\cal E}_R(\tau) {\cal E}_R(\tau') e^{-iE(\tau+\tau')} T \hat\phi(\tau) \hat\phi(\tau') |0_M\rangle+ \int d\chi\int d\chi' {\cal E}_P(\chi) {\cal E}_P(\chi') e^{a(\chi+\chi')} e^{iE(\chi+\chi')} T \hat\phi(\chi) \hat\phi(\chi')|0_M\rangle ,\\
    |A_R\rangle & =
    \int d\tau {\cal E}_R(\tau) e^{-iE\tau} \hat\phi(\tau) |0_M\rangle, \quad 
    |A_P\rangle= \int d\chi {\cal E}_P(\chi) e^{a\chi} e^{iE\chi} \hat\phi(\chi) |0_M\rangle,\\
    |X\rangle & = \int d\tau\int d\chi {\cal E}_R(\tau) {\cal E}_P(\chi) e^{a\chi} e^{-iE(\tau-\chi)} \hat\phi(\tau) \hat\phi(\chi)|0_M\rangle.
\end{aligned}
\end{equation}
Tracing over the field degrees of freedom gives the corresponding two-detector density matrix in the basis $|00\rangle$, $|01\rangle$, $|10\rangle$ and $|11\rangle$
    \begin{equation}
        \rho= \frac{1}{N} \begin{pmatrix}
            \langle1-C|1-C\rangle & 0 & 0 & -\langle X|1-C\rangle \\
            0 & \langle A_R|A_R\rangle & \langle A_P|A_R\rangle & 0 \\
            0 & \langle A_R|A_P\rangle & \langle A_P|A_P\rangle & 0 \\
            -\langle1-C|X\rangle & 0 & 0 & \langle X|X\rangle
        \end{pmatrix},
    \end{equation}
where $N$ is the overall normalization factor. To second order, the negativity ${\cal N}(\rho)$ is approximately~\cite{reznik2005s}
\begin{equation}
    {\cal N}(\rho)\approx \frac{|\langle0_M|X\rangle|- \sqrt{\langle A_R|A_R\rangle\langle A_P|A_P\rangle}}{N},
\end{equation}
where we have
\begin{align}
    \langle A_R|A_R\rangle=& \int d\tau\int d\tau' {\cal E}_R(\tau) {\cal E}_R(\tau') e^{-iE(\tau-\tau')} \langle 0_M|\hat\phi(\tau') \hat\phi(\tau)|0_M\rangle,\\
    \langle A_P|A_P\rangle=& \int d\chi\int d\chi' {\cal E}_P(\chi) {\cal E}_P(\chi') e^{a(\chi+\chi')} e^{-iE(\chi-\chi')} \langle 0_M|\hat\phi(\chi') \hat\phi(\chi)|0_M\rangle.
\end{align}
Choosing the window function for the detector on the right equal to the window function for the detector on the past, we get $\langle A_R|A_R\rangle= \langle A_P|A_P\rangle$. Then, the non-separability condition can be written as $I_X>I_A$.

Ordinary regularized Wightman function in Minkowski spacetime takes the form
\begin{equation}
    \langle0_M|\hat\phi(x) \hat\phi(x')|0_M\rangle= -\frac{1}{(t-t'-i\epsilon)^2-(x-x')^2}.
\end{equation}
Thus for the given trajectories, a coordinate transformation yields (for appropriately rescaled infinitesimal regulator $\epsilon$)
\begin{align}
    \langle0_M|\hat\phi(\tau') \hat\phi(\tau)|0_M\rangle=& \frac{a^2}{4 \sinh^2\left(\frac{a}{2}(\tau-\tau')-i\epsilon\right)}, \quad \langle0_M|\hat\phi(\chi) \hat\phi(\chi')|0_M\rangle= \frac{a^2 e^{-a(\chi+\chi')}}{4 \sinh^2\left(\frac{a}{2}(\chi-\chi')-i\epsilon\right)},\\
    \langle0_M|\hat\phi(\tau) \hat\phi(\chi)|0_M\rangle=& - \frac{a^2 e^{-a\chi}}{2(\sinh a\tau+ \sinh a\chi)}= -\frac{a^2 e^{-a\chi}}{4\cosh \left(a\frac{\chi-\tau}{2}\right) \sinh \left(a\frac{\chi+\tau}{2}- i\epsilon\right)}.
\end{align}
\end{widetext}

To compute $I_X$, we perform following coordinate transformations
\begin{equation}
    {\cal A}= \chi+\tau, \quad {\cal B}= \chi-\tau.
\end{equation}
This gives
\begin{equation}
    \chi= \frac{{\cal A}+{\cal B}}{2}, \quad \tau= \frac{{\cal A}-{\cal B}}{2},
\end{equation}
and $d\chi d\tau= d{\cal A} d{\cal B}/2$. We thus obtain
\begin{align}
    I_X=& \bigg|\frac{1}{2} \int d{\cal A}\int d{\cal B} e^{-({\cal A}^2+{\cal B}^2)/2} e^{iE{\cal B}} \\
    & \frac{1}{\cosh\left(a{\cal B}/2\right) \sinh\left(a{\cal A}/2\right)}\bigg|
    = \bigg|\frac{I_{\cal A} I_{\cal B}}{2}\bigg|,
\end{align}
where
\begin{align}
    I_{\cal A}=& \int d{\cal A} e^{-{\cal A}^2/2} \frac{1}{ \sinh\left(a{\cal A}/2\right)},\\
    I_{\cal B}=& \int d{\cal B} e^{-{\cal B}^2/2} e^{iE{\cal B}} \frac{1}{\cosh\left(a{\cal B}/2\right) }.
\end{align}
The integrand of $I_{\cal A}$ is an odd function and also has a pole at ${\cal A}= 0$. So, the integral $I_{\cal A}$ evaluates to $-i\pi$. $I_{\cal B}$ is non-singular and may also be numerically integrated to obtain $1.387$. We thus obtain $I_X= 2.179$.

A symmetric translation of the Gaussian window functions about the origin $(t,z)=(0,0)$, such that their peaks in the respective past and right Rindler wedges remain null-separated, leaves the integral $I_{\cal B}$ invariant, as it depends solely on the difference of coordinates that are equally translated. Moreover, since the integrand of $I_{\cal A}$ is an odd function, the integral vanishes over the symmetric interval $-\infty$ to $\infty$, except at the pole. The contribution from the pole remains unaffected by the translation of the Gaussian window functions. Consequently, the integral $I_X$ remains invariant under symmetric translation. Similarly, the integral $I_A$ remains invariant, as it also depends solely on the difference of coordinates that are equally translated. Therefore, symmetric translation of the Gaussian window functions about the origin yields a new pair of window functions that leave the sign of the negativity invariant.


\begin{thebibliography}{100}

\bibitem{parker2009}
L.~E.~Parker and D.~Toms,
%``Quantum Field Theory in Curved Spacetime: Quantized Field and Gravity,''
\href{https://doi.org/10.1017/CBO9780511813924}{Cambridge University Press, 2009,
ISBN 978-0-521-87787-9, 978-0-521-87787-9, 978-0-511-60155-2.}

\bibitem{birrell1982}
N.~D.~Birrell and P.~C.~W.~Davies,
%``Quantum Fields in Curved Space,''
\href{https://doi.org/10.1017/CBO9780511622632}{Cambridge University Press, 1982,
ISBN 978-0-521-27858-4, 978-0-521-27858-4}

\bibitem{unruh1976}
W.~G.~Unruh,
%``Notes on black hole evaporation,''
\href{https://doi.org/10.1103/PhysRevD.14.870}{Phys. Rev. D \textbf{14}, 870 (1976).}

\bibitem{crispino2008}
L.~C.~B.~Crispino, A.~Higuchi and G.~E.~A.~Matsas,
%``The Unruh effect and its applications,''
\href{https://doi.org/10.1103/RevModPhys.80.787}{Rev. Mod. Phys. \textbf{80}, 787-838 (2008).} [arXiv:0710.5373 [gr-qc]].

\bibitem[Ginzburg (1987)]{ginzburg1987} V.~L.~Ginzburg and V.~P.~Frolov, \href{https://iopscience.iop.org/article/10.1070/PU1987v030n12ABEH003071/pdf}{Sov. Phys. Usp., 30, 1073 (1987).}

\bibitem[Sciama (1981)]{sciama1981} D.~W.~Sciama, P.~Candelas and D.~Deutsch, \href{https://doi.org/10.1080/00018738100101457}{Advances in Physics, 30(3), 327-366 (1981).}

\bibitem{Martin-Martinez2010}
E.~Martin-Martinez, I.~Fuentes and R.~B.~Mann,
%``Using Berry's phase to detect the Unruh effect at lower accelerations,''
\href{https://doi.org/10.1103/PhysRevLett.107.131301}{Phys. Rev. Lett. \textbf{107}, 131301 (2011)} [arXiv:1012.2208 [quant-ph]].

\bibitem{ispirian2012}
K.~A.~Ispirian,
%``High energy experimental proposals for the study of Unruh (effect) radiation,''
\href{https://web.archive.org/web/20200324092809/http://inspirehep.net/record/1182421/files/article_2012_1_209.pdf}{Prob. Atomic Sci. Technol. \textbf{2012N1}, 209-212 (2012).}

\bibitem{carballo2019unruh}
R.~Carballo-Rubio, L.~J.~Garay, E.~Mart{\'\i}n-Mart{\'\i}nez and J.~De Ram{\'o}n, \href{https://journals.aps.org/prl/abstract/10.1103/PhysRevLett.123.041601}{Phys. Rev. Lett. \textbf{123}, 041601 (2019)}

\bibitem{sudhir2021unruh}
V.~Sudhir, N.~Stritzelberger A.~Kempf,
\href{https://journals.aps.org/prd/abstract/10.1103/PhysRevD.103.105023}{Phys. Rev. D \textbf{103}, 105023 (2021)}

\bibitem{arrechea2021inversion}
J.~Arrechea, C.~Barcel{\'o}, L.~J.~Garay and G.~Garc{\'\i}a-Moreno,
\href{https://journals.aps.org/prd/abstract/10.1103/PhysRevD.104.065004}{Phys. Rev. D \textbf{104}, 065004 (2021)}

\bibitem{quach2021}
J.~Q.~Quach, T.~C.~Ralph and W.~J.~Munro,
%``Berry Phase from the Entanglement of Future and Past Light Cones: Detecting the Timelike Unruh Effect,''
\href{https://doi.org/10.1103/PhysRevLett.129.160401}{Phys. Rev. Lett. \textbf{129}, no.16, 160401 (2022)} [arXiv:2112.00898 [gr-qc]].

\bibitem{davies1974}
P.~C.~W.~Davies,
%``Scalar particle production in Schwarzschild and Rindler metrics,''
\href{https://doi.org/10.1088/0305-4470/8/4/022}{J. Phys. A \textbf{8}, 609-616 (1975).}

\bibitem{Fuentes-Schuller2004} I.~Fuentes-Schuller and R.~B.~Mann,
%``Alice falls into a black hole: Entanglement in non-inertial frames,'' 
\href{https://doi.org/10.1103/PhysRevLett.95.120404}{Phys. Rev. Lett. \textbf{95}, 120404 (2005)} [arXiv:quant-ph/0410172 [quant-ph]].

\bibitem{MTW:73}  C.~Misner, K.~Thorne, and J.~A.~Wheeler, \textit{Gravitation} (Princeton University Press, 1973).

%\bibitem{alsing2003} P.~M.~Alsing and G.~J.~Milburn,
%``Teleportation with a uniformly accelerated partner,'' Phys. Rev. Lett. \textbf{91}, 180404 (2003) doi:10.1103/PhysRevLett.91.180404 [arXiv:quant-ph/0302179 [quant-ph]].

%\bibitem{hu2012} B.~L.~Hu, S.~Y.~Lin and J.~Louko,
%``Relativistic Quantum Information in Detectors-Field Interactions,'' Class. Quant. Grav. \textbf{29}, 224005 (2012) doi:10.1088/0264-9381/29/22/224005 [arXiv:1205.1328 [quant-ph]].

%\bibitem{landulfo2009} A.~G.~S.~Landulfo and G.~E.~A.~Matsas,
%``Sudden death of entanglement and teleportation fidelity loss via the Unruh effect,'' Phys. Rev. A \textbf{80}, 032315 (2009) doi:10.1103/PhysRevA.80.032315 [arXiv:0907.0485 [gr-qc]].

%\bibitem{tjoa2022} E.~Tjoa,
%``Quantum teleportation with relativistic communication from first principles,'' Phys. Rev. A \textbf{106}, no.3, 032432 (2022) doi:10.1103/PhysRevA.106.032432 [arXiv:2206.09294 [quant-ph]].

%\bibitem[Sewell(1982)]{sewell1982} G.~L.~Sewell, Annals of Physics, 141, 201 (1982). doi:10.1016/0003-4916(82)90285-8

%\bibitem{kay1991} B.~S.~Kay and R.~M.~Wald,
%``Theorems on the Uniqueness and Thermal Properties of Stationary, Nonsingular, Quasifree States on Space-Times with a Bifurcate Killing Horizon,'' Phys. Rept. \textbf{207}, 49-136 (1991) doi:10.1016/0370-1573(91)90015-E

%\bibitem[Fulling (1973)]{fulling1973} S.~A.~Fulling, Phys. Rev. D, 7, 2850 (1973). doi:10.1103/PhysRevD.7.2850

\bibitem{olson2011}
S.~J.~Olson and T.~C.~Ralph,
%``Entanglement between the future and past in the quantum vacuum,''
\href{https://doi.org/10.1103/PhysRevLett.106.110404}{Phys. Rev. Lett. \textbf{106}, 110404 (2011).} [arXiv:1003.0720 [quant-ph]].

\bibitem{higuchi2017}
A.~Higuchi, S.~Iso, K.~Ueda and K.~Yamamoto,
%``Entanglement of the Vacuum between Left, Right, Future, and Past: The Origin of Entanglement-Induced Quantum Radiation,''
\href{https://doi.org/10.1103/PhysRevD.96.083531}{Phys. Rev. D \textbf{96}, no.8, 083531 (2017)} [arXiv:1709.05757 [hep-th]].

\bibitem{ueda2021}
K.~Ueda, A.~Higuchi, K.~Yamamoto, A.~Rohim and Y.~Nan,
%``Entanglement of the Vacuum between Left, Right, Future, and Past: Dirac spinor in Rindler spaces and Kasner spaces,''
\href{https://doi.org/10.1103/PhysRevD.103.125005}{Phys. Rev. D \textbf{103}, 125005 (2021)} [arXiv:2104.06625 [gr-qc]].

%\bibitem[Heitler(1954)]{heitler1954} W.~Heitler, International Series of Monographs on Physics, Oxford: Clarendon, 1954, 3rd ed.

\bibitem{reznik2000}
B.~Reznik,
%``Distillation of vacuum entanglement to EPR pairs,''
\href{https://doi.org/10.48550/arXiv.quant-ph/0008006}{[arXiv:quant-ph/0008006 [quant-ph]].}

\bibitem{reznik2005}
B.~Reznik, A.~Retzker and J.~Silman,
%``Violating Bell's inequalities in the vacuum,''
\href{https://doi.org/10.1103/PhysRevA.71.042104}{Phys. Rev. A \textbf{71}, 042104 (2005).} [arXiv:quant-ph/0310058 [quant-ph]].

\bibitem{olson2012}
S.~J.~Olson and T.~C.~Ralph,
%``Extraction of timelike entanglement from the quantum vacuum,''
\href{https://doi.org/10.1103/PhysRevA.85.012306}{Phys. Rev. A \textbf{85}, 012306 (2012).} [arXiv:1101.2565 [quant-ph]].

\bibitem{ralph2014}
T.~C.~Ralph and N.~Walk,
%``Quantum key distribution without sending a quantum signal,''
\href{https://doi.org/10.1088/1367-2630/17/6/063008}{New J. Phys. \textbf{17}, no.6, 063008 (2015).} [arXiv:1409.2061 [quant-ph]].

\bibitem{koga2018}
J.~I.~Koga, G.~Kimura and K.~Maeda,
%``Quantum teleportation in vacuum using only Unruh-DeWitt detectors,''
\href{https://doi.org/10.1103/PhysRevA.97.062338}{Phys. Rev. A \textbf{97}, no.6, 062338 (2018).} [arXiv:1804.01183 [gr-qc]].

\bibitem{Ng2018}
K.~K.~Ng, R.~B.~Mann and E.~Mart{\'\i}n-Mart{\'\i}nez,
%``New techniques for entanglement harvesting in flat and curved spacetimes,''
\href{https://doi.org/10.1103/PhysRevD.97.125011}{Phys. Rev. D \textbf{97}, no.12, 125011 (2018).} [arXiv:1805.01096 [quant-ph]].

\bibitem{Martin-Martinez:2015}
E.~Martin-Martinez and B.~C.~Sanders,
%``Precise space\textendash{}time positioning for entanglement harvesting,''
\href{https://doi.org/10.1088/1367-2630/18/4/043031}{New J. Phys. \textbf{18}, 043031 (2016).} [arXiv:1508.01209 [quant-ph]].

\bibitem{lin2015}
S.~Y.~Lin, K.~Shiokawa, C.~H.~Chou and B.~L.~Hu,
%``Quantum teleportation between moving detectors,''
\href{https://doi.org/10.1103/PhysRevD.91.084063}{Phys. Rev. D \textbf{91}, no.8, 084063 (2015).} [arXiv:1502.03539 [gr-qc]].

\bibitem{landulfo2009}
A.~G.~S.~Landulfo and G.~E.~A.~Matsas,
%``Sudden death of entanglement and teleportation fidelity loss via the Unruh effect,''
\href{https://doi.org/10.1103/PhysRevA.80.032315}{Phys. Rev. A \textbf{80}, 032315 (2009).} [arXiv:0907.0485 [gr-qc]].

\bibitem{foo2020}
J.~Foo and T.~C.~Ralph,
%``Continuous-variable quantum teleportation with vacuum-entangled Rindler modes,''
\href{https://doi.org/10.1103/PhysRevD.101.085006}{Phys. Rev. D \textbf{101}, no.8, 085006 (2020).} [arXiv:2001.03387 [quant-ph]].

\bibitem{koch2007}
J.~Koch, T.~M.~Yu, J.~Gambetta, A.~A.~Houck, D.~I.~Schuster, J.~Majer, A.~Blais, M.~H.~Devoret, S.~M.~Girvin and R.~J.~Schoelkopf,
%``Charge-insensitive qubit design derived from the Cooper pair box,''
\href{https://doi.org/10.1103/physreva.76.042319}{Phys. Rev. A \textbf{76}, no.4, 042319 (2007).} [arXiv:cond-mat/0703002 [cond-mat.mes-hall]].

\bibitem{zhu2021} D. Zhu, T. Jaako, Q. He, \etal, \href{https://doi.org/10.1103/PhysRevApplied.16.014024}{Physical Review Applied, 16, 1, 014024 (2021).}

\bibitem[Barrett et al.(2005)]{barrett2005} J.~Barrett, L.~Hardy, \& A.~Kent, \href{https://doi.org/10.1103/PhysRevLett.95.010503}{Phys. Rev. Lett., No Signaling and Quantum Key Distribution, 95, 1, 010503 (2005).}


\bibitem[Rindler(1966)]{rindler1966} W.~Rindler, \href{https://doi.org/10.1119/1.1972547}{American Journal of Physics, Kruskal Space and the Uniformly Accelerated Frame, 34, 12, 1174 (1966).}

\bibitem{barcelo2011}
C.~Barcelo, S.~Liberati, S.~Sonego and M.~Visser,
%``Hawking-like radiation from evolving black holes and compact horizonless objects,''
\href{https://doi.org/10.1007/JHEP02(2011)003}{JHEP \textbf{02}, 003 (2011)}.
[arXiv:1011.5911 [gr-qc]].

%\bibitem{rosu2001} H.~C.~Rosu,
%``Hawking like effects and Unruh like effects: Toward experiments?,'' Grav. Cosmol. \textbf{7}, 1-17 (2001) [arXiv:gr-qc/9406012 [gr-qc]].

%\bibitem{nikishov1988} A.~I.~Nikishov and V.~I.~Ritus,
%``Processes Induced by a Charged Particle in an Electric Field, and the Unruh Heat Bath Concept,'' \href{http://www.jetp.ras.ru/cgi-bin/e/index/r/94/7/p31?a=list}{Sov. Phys. JETP \textbf{67}, 1313-1321 (1988).}

%\bibitem{padmanabhan1990} T.~Padmanabhan,
%``Physical interpretation of quantum field theory in noninertial coordinate systems,'' Phys. Rev. Lett. \textbf{64}, 2471-2474 (1990) doi:10.1103/PhysRevLett.64.2471

\bibitem{gerlach1988}
U.~H.~Gerlach,
%``MINKOWSKI BESSEL MODES,''
\href{https://doi.org/10.1103/PhysRevD.38.514}{Phys. Rev. D \textbf{38}, 514-521 (1988).} [arXiv:gr-qc/9910097 [gr-qc]].

\bibitem{takagi1986}
S.~Takagi,
%``Vacuum Noise and Stress Induced by Uniform Acceleration: Hawking-Unruh Effect in Rindler Manifold of Arbitrary Dimension,''
\href{https://doi.org/10.1143/PTP.88.1}{Prog. Theor. Phys. Suppl. \textbf{88}, 1-142 (1986).}

\bibitem[Fulling et al.(1974)]{fulling1974} S.~A.~Fulling, L.~Parker, \& B.~L.~Hu, \href{https://doi.org/10.1103/PhysRevD.10.3905}{Phys. Rev. D, 10, 3905 (1974).}

\bibitem[Carroll(2004)]{carroll2004} S.~M. Carroll, \ 2004, Spacetime and geometry. San Francisco, CA, USA: Addison Wesley, ISBN 0-8053-8732-3, 2004, XIV + 513 pp.

%\bibitem{ispirian2012} K.~A.~Ispirian,
%``High energy experimental proposals for the study of Unruh (effect) radiation,'' Prob. Atomic Sci. Technol. \textbf{2012N1}, 209-212 (2012)

%\bibitem{Martin-Martinez2010} E.~Martin-Martinez, I.~Fuentes and R.~B.~Mann,
%``Using Berry's phase to detect the Unruh effect at lower accelerations,'' Phys. Rev. Lett. \textbf{107}, 131301 (2011) doi:10.1103/PhysRevLett.107.131301 [arXiv:1012.2208 [quant-ph]].

%\bibitem{quach2021} J.~Q.~Quach, T.~C.~Ralph and W.~J.~Munro,
%``Berry Phase from the Entanglement of Future and Past Light Cones: Detecting the Timelike Unruh Effect,'' Phys. Rev. Lett. \textbf{129}, no.16, 160401 (2022) doi:10.1103/PhysRevLett.129.160401 [arXiv:2112.00898 [gr-qc]].

\bibitem{krantz2019}
P.~Krantz, M.~Kjaergaard, F.~Yan, T.~P.~Orlando, S.~Gustavsson and W.~D.~Oliver,
%``A quantum engineer's guide to superconducting qubits''
\href{https://doi.org/10.1063/1.5089550}{Appl. Phys. Rev. \textbf{6}, 021318 (2019).}

\bibitem{peres1996}
A.~Peres,
%``Separability criterion for density matrices,''
\href{https://doi.org/10.1103/PhysRevLett.77.1413}{Phys. Rev. Lett. \textbf{77}, 1413-1415 (1996).} [arXiv:quant-ph/9604005 [quant-ph]].

\bibitem{horodecki1996}
M.~Horodecki, P.~Horodecki and R.~Horodecki,
%``On the necessary and sufficient conditions for separability of mixed quantum states,''
\href{https://doi.org/10.1016/S0375-9601(96)00706-2}{Phys. Lett. A \textbf{223}, 1 (1996).} [arXiv:quant-ph/9605038 [quant-ph]].

\bibitem{bennett1992}
C.~H.~Bennett, G.~Brassard, C.~Crepeau, R.~Jozsa, A.~Peres and W.~K.~Wootters,
%``Teleporting an unknown quantum state via dual classical and Einstein-Podolsky-Rosen channels,''
\href{https://doi.org/10.1103/PhysRevLett.70.1895}{Phys. Rev. Lett. \textbf{70}, 1895-1899 (1993).}

\bibitem{wang2020controllable}
Z.~Wang et al., 
\href{https://journals.aps.org/prl/pdf/10.1103/PhysRevLett.124.013601}{Phys. Rev. Lett. \textbf{124}, 013601 (2020)}

\bibitem{reznik2002}
B.~Reznik,
%``Entanglement from the vacuum,''
\href{https://doi.org/10.1023/A:1022875910744}{Found. Phys. \textbf{33}, 167-176 (2003).} [arXiv:quant-ph/0212044 [quant-ph]].

\bibitem{bradler2007}
K.~Br\'adler,
%``Eavesdropping of quantum communication from a noninertial frame,''
\href{https://doi.org/10.1103/PhysRevA.75.022311}{Phys. Rev. A \textbf{75}, 022311 (2007).} [arXiv:quant-ph/0212044 [quant-ph]].

\end{thebibliography}

\begin{thebibliography}{100}
\makeatletter
\providecommand \@ifxundefined [1]{%
 \@ifx{#1\undefined}
}%
\providecommand \@ifnum [1]{%
 \ifnum #1\expandafter \@firstoftwo
 \else \expandafter \@secondoftwo
 \fi
}%
\providecommand \@ifx [1]{%
 \ifx #1\expandafter \@firstoftwo
 \else \expandafter \@secondoftwo
 \fi
}%
\providecommand \natexlab [1]{#1}%
\providecommand \enquote  [1]{``#1''}%
\providecommand \bibnamefont  [1]{#1}%
\providecommand \bibfnamefont [1]{#1}%
\providecommand \citenamefont [1]{#1}%
\providecommand \href@noop [0]{\@secondoftwo}%
\providecommand \href [0]{\begingroup \@sanitize@url \@href}%
\providecommand \@href[1]{\@@startlink{#1}\@@href}%
\providecommand \@@href[1]{\endgroup#1\@@endlink}%
\providecommand \@sanitize@url [0]{\catcode `\\12\catcode `\$12\catcode `\&12\catcode `\#12\catcode `\^12\catcode `\_12\catcode `\%12\relax}%
\providecommand \@@startlink[1]{}%
\providecommand \@@endlink[0]{}%
\providecommand \url  [0]{\begingroup\@sanitize@url \@url }%
\providecommand \@url [1]{\endgroup\@href {#1}{\urlprefix }}%
\providecommand \urlprefix  [0]{URL }%
\providecommand \Eprint [0]{\href }%
\providecommand \doibase [0]{https://doi.org/}%
\providecommand \selectlanguage [0]{\@gobble}%
\providecommand \bibinfo  [0]{\@secondoftwo}%
\providecommand \bibfield  [0]{\@secondoftwo}%
\providecommand \translation [1]{[#1]}%
\providecommand \BibitemOpen [0]{}%
\providecommand \bibitemStop [0]{}%
\providecommand \bibitemNoStop [0]{.\EOS\space}%
\providecommand \EOS [0]{\spacefactor3000\relax}%
\providecommand \BibitemShut  [1]{\csname bibitem#1\endcsname}%
\let\auto@bib@innerbib\@empty


\bibitem{crispino2008s}
L.~C.~B.~Crispino, A.~Higuchi and G.~E.~A.~Matsas,
%``The Unruh effect and its applications,''
\href{https://doi.org/10.1103/RevModPhys.80.787}{Rev. Mod. Phys. \textbf{80}, 787-838 (2008).} [arXiv:0710.5373 [gr-qc]].

\bibitem{takagi1986s}
S.~Takagi,
%``Vacuum Noise and Stress Induced by Uniform Acceleration: Hawking-Unruh Effect in Rindler Manifold of Arbitrary Dimension,''
\href{https://doi.org/10.1143/PTP.88.1}{Prog. Theor. Phys. Suppl. \textbf{88}, 1-142 (1986).}

\bibitem{olson2011s}
S.~J.~Olson and T.~C.~Ralph,
%``Entanglement between the future and past in the quantum vacuum,''
\href{https://doi.org/10.1103/PhysRevLett.106.110404}{Phys. Rev. Lett. \textbf{106}, 110404 (2011).} [arXiv:1003.0720 [quant-ph]].

\bibitem{reznik2005s}
B.~Reznik, A.~Retzker and J.~Silman,
%``Violating Bell's inequalities in the vacuum,''
\href{https://doi.org/10.1103/PhysRevA.71.042104}{Phys. Rev. A \textbf{71}, 042104 (2005).} [arXiv:quant-ph/0310058 [quant-ph]].

\end{thebibliography}
\end{document}